\documentstyle[13lomcon,cite,epsfig]{article}
\newcommand{\beq}{\begin{equation}}
\newcommand{\eeq}{\end{equation}}
\newcommand{\preprint}{\newline%
  \begin{picture}(400,0)
  \put(250,100){\rm\small HU--EP--08/01, ITEP-LAT/2008-08}
  \end{picture}}

\begin{document}
\title{LATTICE  RESULTS ON GLUON AND GHOST PROPAGATORS IN LANDAU GAUGE 
\preprint
}

\author{I.L.~Bogolubsky}
\address{Joint Institute for Nuclear Research, 141980 Dubna, Russia}

\author{V.G.~Bornyakov \footnote{e-mail: vitaly.bornyakov@ihep.ru}}
\address{Institute for High Energy Physics, 142281 Protvino, Russia and \\ 
Institute of Theoretical and Experimental Physics,
Moscow, Russia}

\author{G.~Burgio}
\address{Universit\"at T\"ubingen, Institut f\"ur Theoretische Physik,
72076 T\"ubingen, Germany}

\author{E.-M.~Ilgenfritz,
M.~M\"uller--Preussker}
\address{Humboldt-Universit\"at zu Berlin, Institut f\"ur Physik,
12489 Berlin, Germany}

\author{ V.K.~Mitrjushkin} 
\address{Joint Institute for Nuclear Research, 141980 Dubna, Russia and \\ 
Institute of Theoretical and Experimental Physics, Moscow, Russia}

\maketitle

\abstracts{ 
We present clear evidence of strong effects of Gribov copies in
Landau gauge gluon and ghost propagators computed on the lattice
at small momenta by employing 
a new approach to Landau gauge fixing and a more effective numerical algorithm. 
It is further shown that
the new approach substantially decreases notorious finite-volume effects.}

\section{Introduction}
The gauge-variant Green functions, in particular for the covariant 
Landau gauge, are important for various reasons. 
Their infrared asymptotics is crucial for gluon and quark confinement
according to scenarios invented by Gribov~\cite{Gribov:1977wm} and 
Zwanziger~\cite{Zwanziger:1993dh} and by Kugo and Ojima~\cite{Kugo:1979gm}. 
They have proposed that the Landau gauge ghost propagator 
is infrared diverging while the gluon propagator is infrared vanishing. 
The interest in these propagators was stimulated in part by the progress 
achieved in solving Dyson-Schwinger equations (DSE) for these propagators 
(for a recent review see~\cite{Alkofer:2006jf}).
Recently it has been argued that a unique and exact power-like infrared 
asymptotic behavior of all Green functions can be derived without truncating 
the hierarchy of DSE~\cite{Fischer:2006vf}. This solution agrees completely 
with the scenarios of confinement mentioned above. The lattice approach is 
another powerful tool to compute these propagators in an {\it ab initio}
fashion but not free of lattice artefacts. So far, there is no consensus
between DSE and lattice results. For the gluon propagator, the ultimate 
decrease towards  vanishing momentum has not yet been established in lattice 
computations. Lattice results for the ghost propagator qualitatively agree 
with the predicted diverging behavior but show a substantially smaller
infrared exponent~\cite{Ilgenfritz:2006he}.

The lattice approach has its own limitations. The effects of the finite 
volume might be strong at the lowest lattice momenta. Moreover, gauge 
fixing is not unique resulting in the so-called {\it Gribov problem}.
Previously it has been concluded that the gluon propagator does not show
effects of Gribov copies beyond statistical noise,
while the ghost propagator has been found to deviate by up to 10\%
depending on the quality of gauge 
fixing~\cite{Bakeev:2003rr,Sternbeck:2005tk}.

Recently a new, extended approach to Landau gauge fixing has been 
proposed~\cite{Bogolubsky:2005wf}. 
In this contribution we present results obtained within this new method
and using a more effective numerical algorithm for lattice gauge fixing, 
the simulated annealing (SA) algorithm. Results for the gluon propagator 
have been already discussed in~\cite{Bogolubsky:2007bw}, while results 
for the ghost propagator are presented here for the first time. 

\section{Computational details}
Our computations have been performed for one lattice spacing corresponding
to rather strong bare coupling, at $\beta \equiv 4/g_0^2 = 2.20$, on lattices 
from $8^4$ up to $32^4$. The corresponding lattice scale $a$ is fixed 
adopting $\sqrt{\sigma} a = 0.469$ \cite{Fingberg:1992ju} 
with the string tension put equal to $\sigma$ = (440 MeV)$^2$. 
Thus, our largest lattice size $32^4$ corresponds to a volume 
$(6.7 {\rm~fm})^4$.

In order to fix the Landau gauge for each lattice gauge field
$\{U\}$ generated by means of a MC procedure, the gauge functional
\beq
F[g]= \frac{1}{2 N_{links}} \sum_{x,\mu}
      \mathrm{tr} \left( g(x) U_{x\mu} g^{\dagger}(x+\hat{\mu}) \right)
\label{gauge_functional}
\eeq
is iteratively maximized with respect to a gauge transformation $~g(x)~$
which is usually taken as a periodic field.
In $SU(N)$ gluodynamics the lattice action and the path integral measure are 
invariant under extended gauge transformations which are 
periodic modulo $Z(N)$,
\beq
g(x+L\hat{\nu}) = z_{\nu} g(x)\,, \qquad z_{\nu} \in Z(N) 
\label{flip_trans}
\eeq
in all four directions.
Any such gauge transformation is equivalent to a combination of a periodic 
gauge transformation and a flip $~U_{x\nu} \to z_{\nu} ~U_{x\nu}~$ for a 3D 
hyperplane with fixed $x_{\nu}$. 
With respect to the flip transformation all gauge copies of one given
field configuration can be split into $N^4$ flip sectors. 
The traditional gauge fixing procedure considers one flip sector as a 
separate gauge orbit. The new approach suggested in \cite{Bogolubsky:2005wf} 
combines all $N^4$ sectors into one gauge orbit. Note, that this approach 
is not applicable in a gauge theory with fundamental matter fields because 
the action is not invariant under transformation (\ref{flip_trans}), while 
in the deconfinement phase of $SU(N)$ pure gluodynamics it should be modified: 
only flips in space directions are left in the gauge orbit.
In practice, few Gribov copies 
are generated for each sector and the best one over all sectors is chosen
by employing an optimized simulating annealing algorithm in combination with
finalizing overrelaxation.

\section{Results}
Thus, we are looking for the gauge copy with the highest value of the gauge functional 
among gauge copies belonging to the enlarged gauge orbit as defined above. 
It is immediately clear that this procedure allows to find higher local maxima
of the gauge functional (\ref{gauge_functional}) than the traditional (`old')
gauge fixing procedures 
employing purely periodic gauge transformations and
the standard overrelaxation algorithm.
Obviously the two prescriptions to fix the Landau gauge, the traditional one
and the extended one, are not equivalent. Indeed, for some modest lattice
volumes and for the lowest momenta it has been shown in Ref.~\cite{Bogolubsky:2005wf}
that they give rise to different results for the gluon as well as the ghost
propagators. Comparing results for different lattice sizes we found 
that the results seem to converge to each other in the large volume limit.
It is important that results obtained with the new prescription converge 
towards the infinite volume limit much faster.
\begin{figure}[hpbt]
\centering
\hspace*{1cm}\epsfxsize=11.truecm\epsfbox{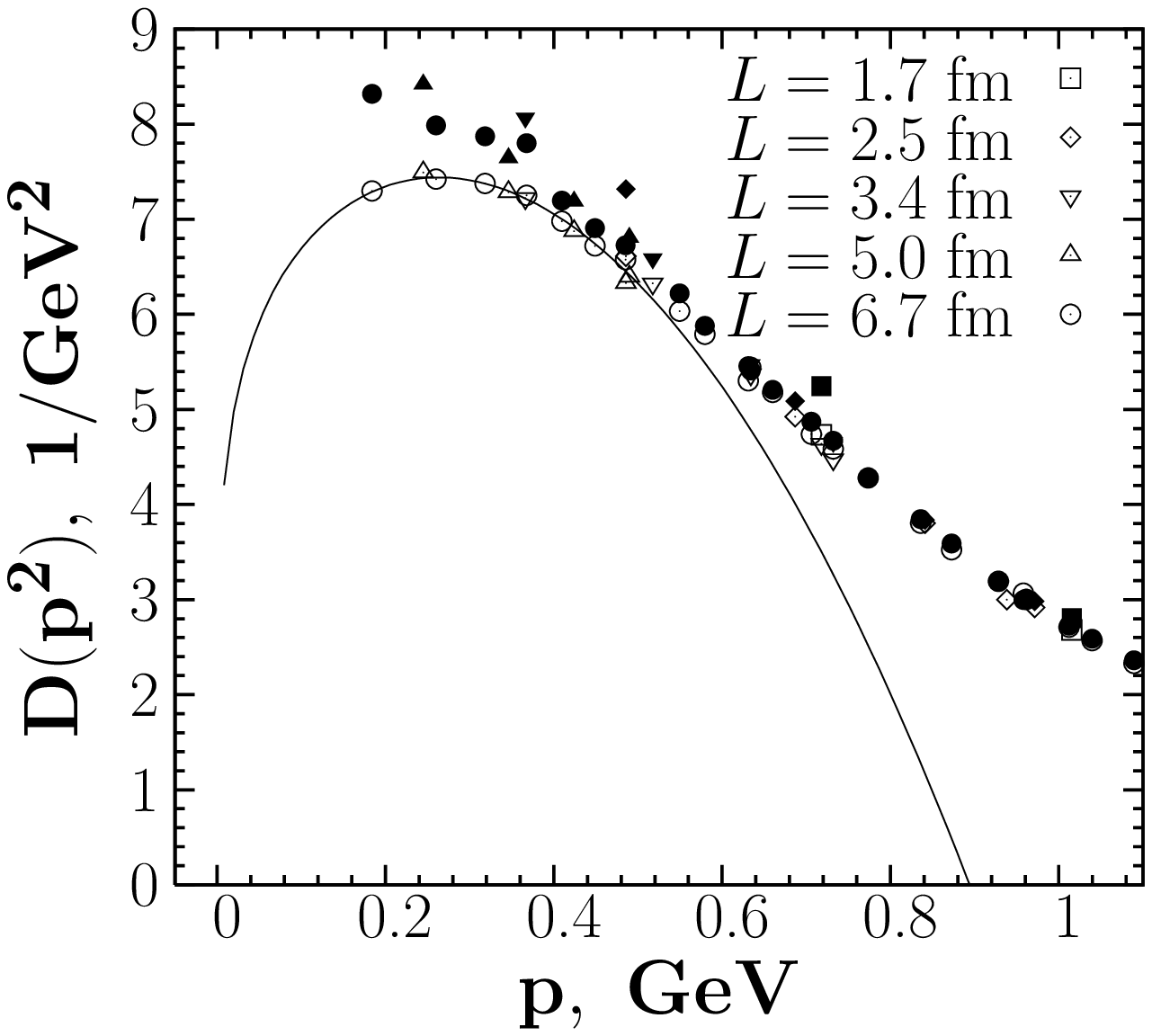}
\caption{Comparison of the gluon propagator computed with new (empty symbols) 
and old (filled symbols) procedures on lattices of various sizes. Error bars 
which are in most cases smaller than symbol sizes are not shown for clarity. The curve
corresponds to eq.~(\ref{fit}). }
\label{glue_prop}
\end{figure}
In Fig.~\ref{glue_prop} the gluon propagator $D(p^2)$ is 
shown.~\footnote{For definition of $D(p^2)$ as well as of the ghost dressing 
function $J(p^2)$ see  Refs.~\cite{Bogolubsky:2005wf,Bogolubsky:2007bw}.}
One can see that the Gribov copy effects are strong up to $p \sim 0.6$ GeV. 
Furthermore, results obtained with the new procedure show no finite volume 
effects while these effects are clearly seen for results obtained with the 
old procedure. We have also checked, whether our result can be seen in 
agreement with the expectation  $~D(p~\to~0)=0$. We made a fit for 
$ 0 < p < 0.5$ GeV  with the function
\beq
D(p^2)=(p^2)^{\,\alpha} \cdot (g_0 + g_1 \cdot p^2 ) \,.
\label{fit}
\eeq
and found $\alpha = 0.09(1)$ which is in qualitative agreement with the DSE 
result \cite{Alkofer:2006jf}. 

\begin{figure}[hpbt]
\centering
\hspace*{1cm}\epsfxsize=11.truecm\epsfbox{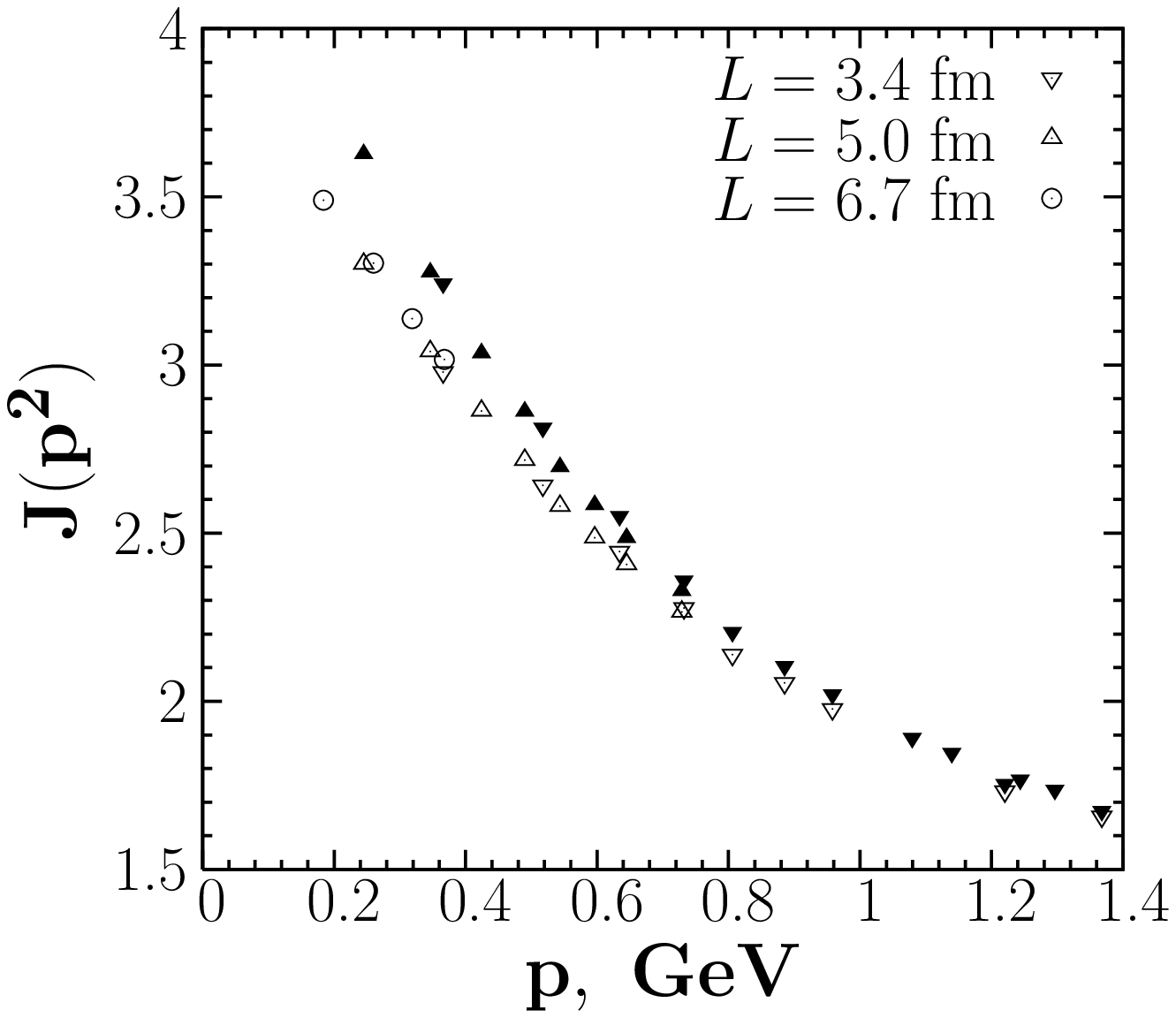}
\caption{Ghost dressing function for two procedures. 
 Symbols are as in Fig.~\ref{glue_prop}.}
\label{ghost_prop}
\end{figure}

In Fig.~\ref{ghost_prop} we show our results for the ghost dressing function 
$J(p^2)$. Again, strong Gribov copies effects are seen. On the other hand, 
the finite-volume effects are not strong for both procedures. We confirm
earlier $SU(3)$ 
results \cite{Ilgenfritz:2006he} that the lattice ghost propagator 
is much less diverging in the infrared limit than it is predicted by DSE 
solutions \cite{Alkofer:2006jf}. Moreover, within the new procedure this 
disagreement becomes even stronger. This is one of the problems to be 
resolved in future studies.

\end{document}